\documentclass[12pt,preprint]{aastex}

\newcommand\plottwovert[2]{\centering \leavevmode
 \includegraphics[width={0.8\columnwidth}]{#1} \hfil
 \includegraphics[width={0.8\columnwidth}]{#2}}

\newcommand\plotthree[3]{\centering \leavevmode
 \includegraphics[width={0.5\columnwidth}]{#1} \hfil
 \includegraphics[width={0.5\columnwidth}]{#2} \hfil
\includegraphics[width={0.5\columnwidth}]{#3}}

\begin{document}

\title{Simulations of Buoyant Bubbles in Galaxy Clusters}

\author{M.Br\"{u}ggen\altaffilmark{1,2}}
\altaffiltext{1}{International University Bremen, Campus Ring 1, 28759 Bremen, Germany}
\altaffiltext{2}{Max-Planck Institut f\"ur Astrophysik,
Karl-Schwarzschild-Str 1, 85740 Garching, Germany}

\email{m.brueggen@iu-bremen.de}

\begin{abstract}
It is generally argued that most clusters of galaxies host cooling
flows in which radiative cooling in the centre causes a slow
inflow. However, recent observations by Chandra and XMM conflict with
the predicted cooling flow rates. Here we report highly resolved
hydrodynamic simulations which show that buoyant bubbles can offset
the cooling in the inner regions of clusters and can significantly
delay the deposition of cold gas. The subsonic rise of bubbles uplifts
colder material from the central regions of the cluster. This colder
material appears as bright rims around the bubbles. The bubbles
themselves appear as depressions in the X-ray surface brightness as
observed in a growing number of clusters.

\end{abstract}




\section{Introduction}

The X-ray surface brightness of many clusters of galaxies shows a
strong central peak which is generally interpreted as the signature of
a cooling flow (Cowie \& Binney 1977, Fabian \& Nulsen 1977, Sarazin
1988, Fabian 1994). However, the simple cooling flow model conflicts
with a growing number of observations that show that while the
temperature is declining in the central region, gas with a temperature
below $\sim$1 keV is significantly less abundant than predicted
(\citealt{pete01}, \citealt{tamu01}). It has been shown that star
formation rates as inferred from observations only amount to a few
percent of the predicted mass deposition rates of 1 - 1000
$M_{\odot}$/yr (e.g. \citealt{mcnam97}, \citealt{craw99}). Moreover,
spectral evidence gathered by XMM-Newton has shown inconsistencies of
the observed spectra with those predicted by cooling flow models
(\citealt{boe02}).\\

Recently, two main candidates for the heating of the gas in the
central regions of clusters have emerged: (i) heating by outflows from
active galactic nuclei (AGN) (Binney \& Tabor 1995, Tabor \& Binney
1993, Churazov et al. 2001, Br\"uggen et al. 2002, \citealt{Bru02},
\citealt{Bas02}, \citealt{RHB01}), and (ii) transport of heat from the
outer regions of the cluster by thermal conduction
(\citealt{zaknar02}, \citealt{nara01}, Fabian, Voigt \& Morris 2002,
\citealt{ruszkow}, \citealt{voig02}, \citealt{gruz02},
\citealt{bert86}). A third possibility that has been discussed is
heating by magnetic fields through reconnection
(e.g. \citealt{sok90}).\\

In this paper we will consider the first scenario. Radio-loud active
galactic nuclei (AGN) drive strong outflows in the form of jets that
inflate bubbles or lobes. The lobes are filled with hot plasma, and
can heat the cluster gas in various ways. In the case of very
energetic jets, the expansion of the lobes will be supersonic, and the
resulting strong shock will heat and compress the gas. However, there
is mounting evidence that weaker jets, which are presumably much more
common, do not lead to efficient shock heating. Nevertheless, they
still produce pockets of low-density gas that displace the hot
intra-cluster medium (ICM) and thus appear as depressions in the X-ray
surface brightness. High-resolution X-ray observations of cooling flow
clusters with Chandra have revealed a multitude of X-ray holes on
scales $<$ 50 kpc, often coincident with patches of radio emission,
e.g. in Abell 2052 (Blanton et al. 2001), Perseus (B\"ohringer et
al. 1993, Churazov et al. 2000, Fabian et al. 2000), Cygnus A
(\citealt{wil00} and \citealt{car94}), Abell 2597 (\citealt{mcnam01}),
MKW3s (\citealt{mazz02}), RBS797 (\citealt{sch01}), HCG62
(\citealt{vrt}) and Centaurus A (\citealt{sax01}). In addition to
radio-bright cavities, radio-faint ``ghost'' cavities have been found,
e.g. in Perseus, which presumably date from previous radio episodes of
the AGN. The pressure support in the cavities is not well understood. In Abell
2052 the external gas pressure is about 10 times higher than the
internal equipartition pressure. The cavities found in MKW3s, however,
are filled with hot thermal gas that provides pressure support against
the ICM and the radio observations of Hydra A also suggest that the
bubbles are in local pressure equilibrium with the surroundings. The
ubiquity of cavities in clusters means that they must be relatively
long-lived and that the majority must be in rough pressure balance
with their surroundings. However, the source of this pressure, the
structure of the magnetic fields inside the bubbles and the equation
of state of the hot plasma are still uncertain.\\

Chandra images of Hydra A, Perseus and Abell 2052 show soft X-ray
emission form the rims of the bubbles which imply that the gas around
the rims of these bubbles is colder and not hotter than the ambient
gas (e.g. Fabian et al. 2000, 2001, \citealt{nul02}). The existence of
these cold rims is taken as evidence that the gas has not been shocked
during the inflation of the bubbles but that the cold rims consist of
low-entropy gas that has been displaced from the centre of the cluster
(Brighenti \& Mathews 2002). Therefore, the radio lobes must have
expanded gently into the ICM. The buoyant rise and the subsequent
mixing of high-entropy gas with the cluster material can lead to a
significant re-structuring of the inner regions of a cooling flow
(e.g., \citealt{Bru02}, \citealt{Bas02}). \\

Assuming that a typical central galaxy produces 10-100 bubbles in its
life time, each carrying an energy of up to $10^{59}$ ergs, the total
energy in bubbles can be comparable to the thermal energy of the X-ray
emitting plasma in the inner cluster regions
(\citealt{mcnam02}). Moreover, there is evidence that the total power
in radio jets is substantially larger than the radio power of the
lobes that they feed (\citealt{bic96}). Thus the question arises of
where the energy ultimately goes and what fraction of the energy is
deposited into the ICM. This is the question that is addressed in this
paper.\\

Both, the rise and the mixing are complex processes that are difficult
to treat on the basis of analytical estimates alone. We therefore
performed highly resolved two-dimensional hydrodynamic simulations of
buoyant gas in a typical cluster environment similar to previous
simulations of lower resolution. Only two-dimensional simulations were
performed because our primary aim was to achieve a very high spatial
resolution in order to minimize the effects of numerical diffusivity
that have hampered previous simulations. This was also aided by the
use of an adaptive grid that places resolution elements only where
they are needed, such that an effective resolution of 2048 times 2048
elements could be achieved. Experience has shown that large-scale
transport properties are not very different in two dimensional and
three-dimensional simulations, so we expect our conclusions to hold
even for the three-dimensional reality.\\

In this paper, we will extend the work presented in
\cite{Bru02}. Apart from a much more detailed analysis, we have
extended previous work by simulating an ambipolar jet in the entire
cluster compared to merely a section of the cluster. Thus, any
artificial surfaces of symmetry are avoided (as also in
\citealt{RHB01}; \citealt{Bas02}) and the amount of global,
distributed heating can be studied. Moreover, we ran the simulations
for a longer physical time, namely for much longer than the central
cooling time of the cluster. We also included radiative cooling and
treated the radio plasma and the ambient plasma as separate fluids
with different equations of state.

\section{Model}

\begin{figure}[htp]
\plotthree{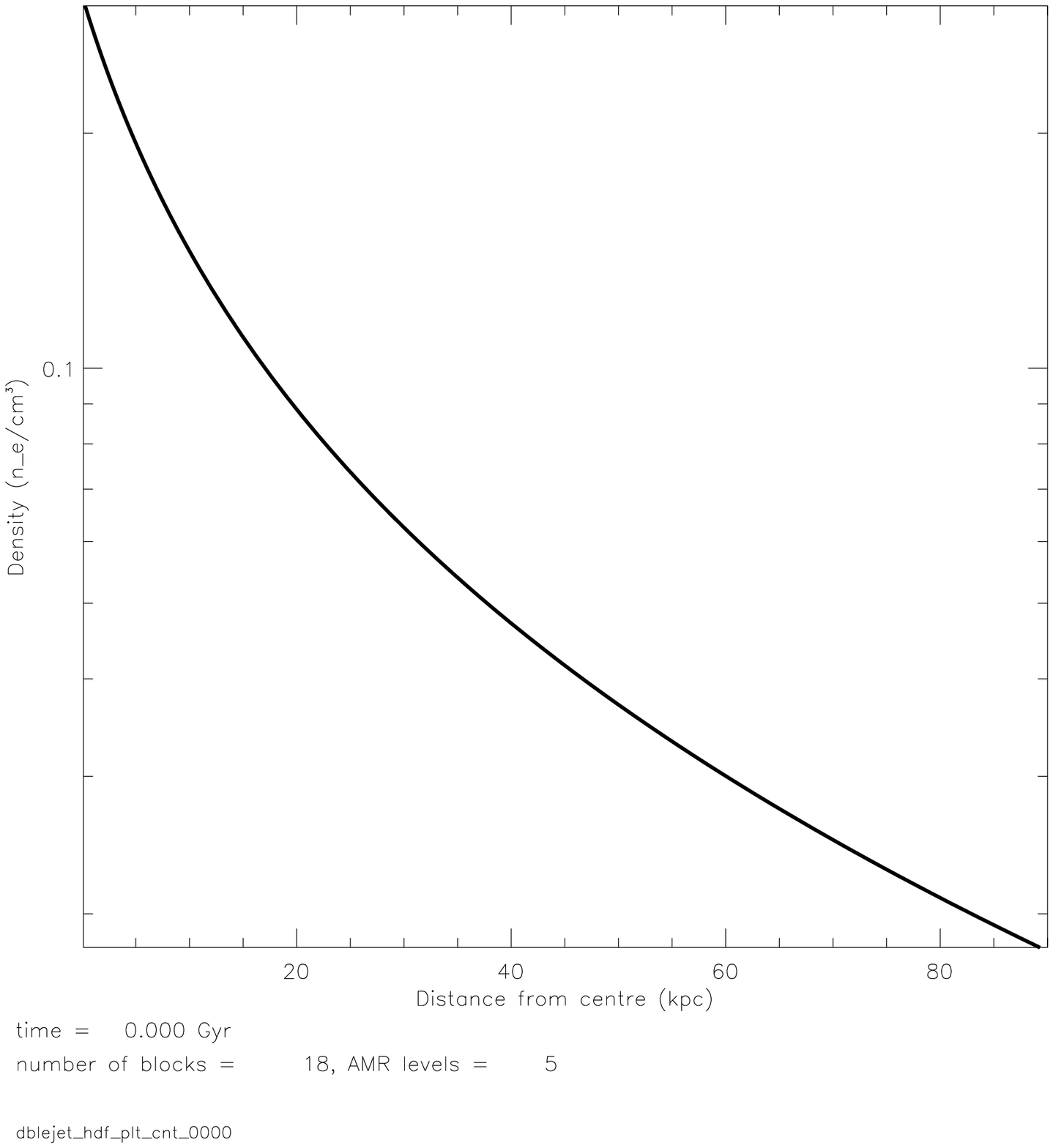}{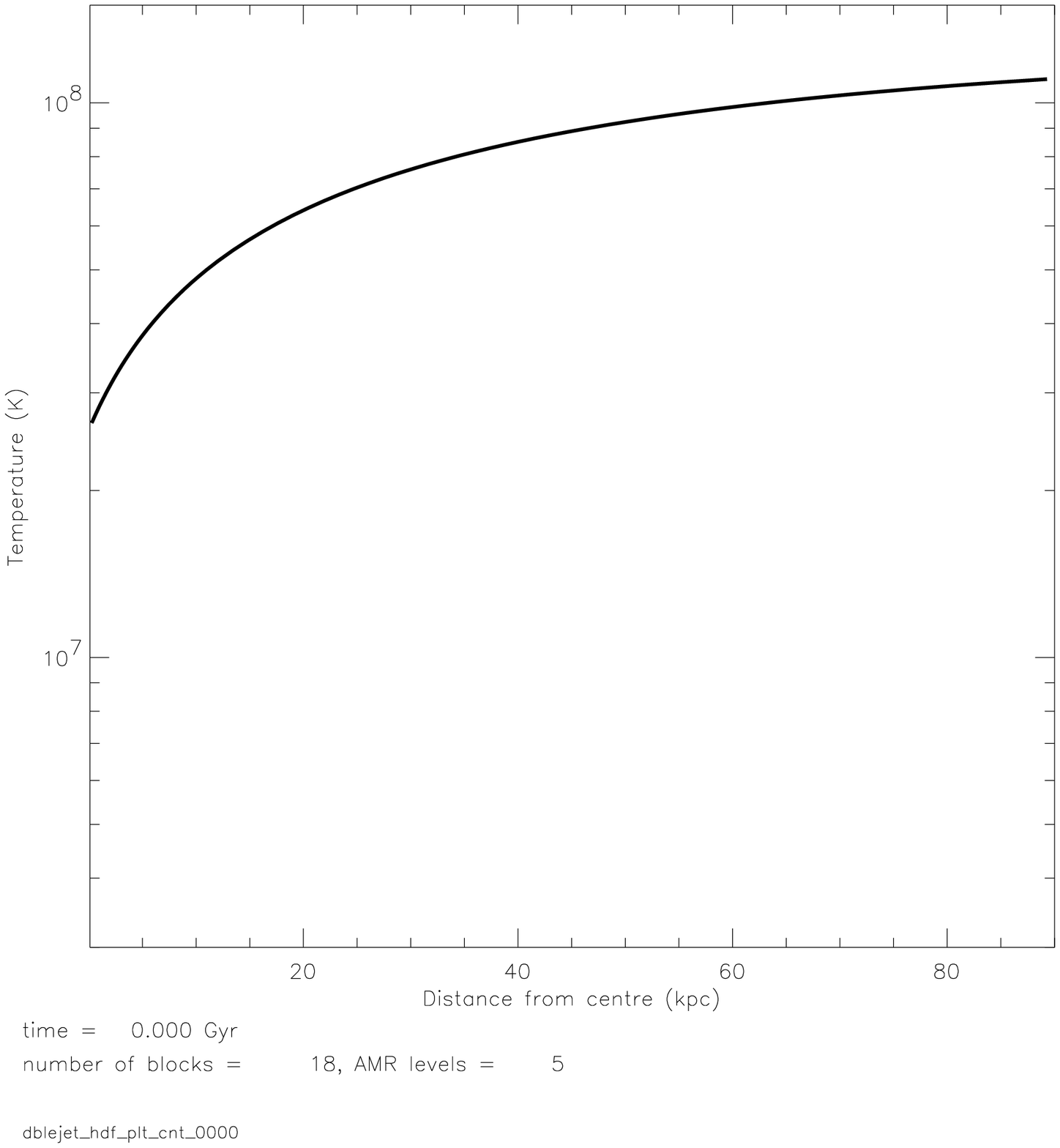}{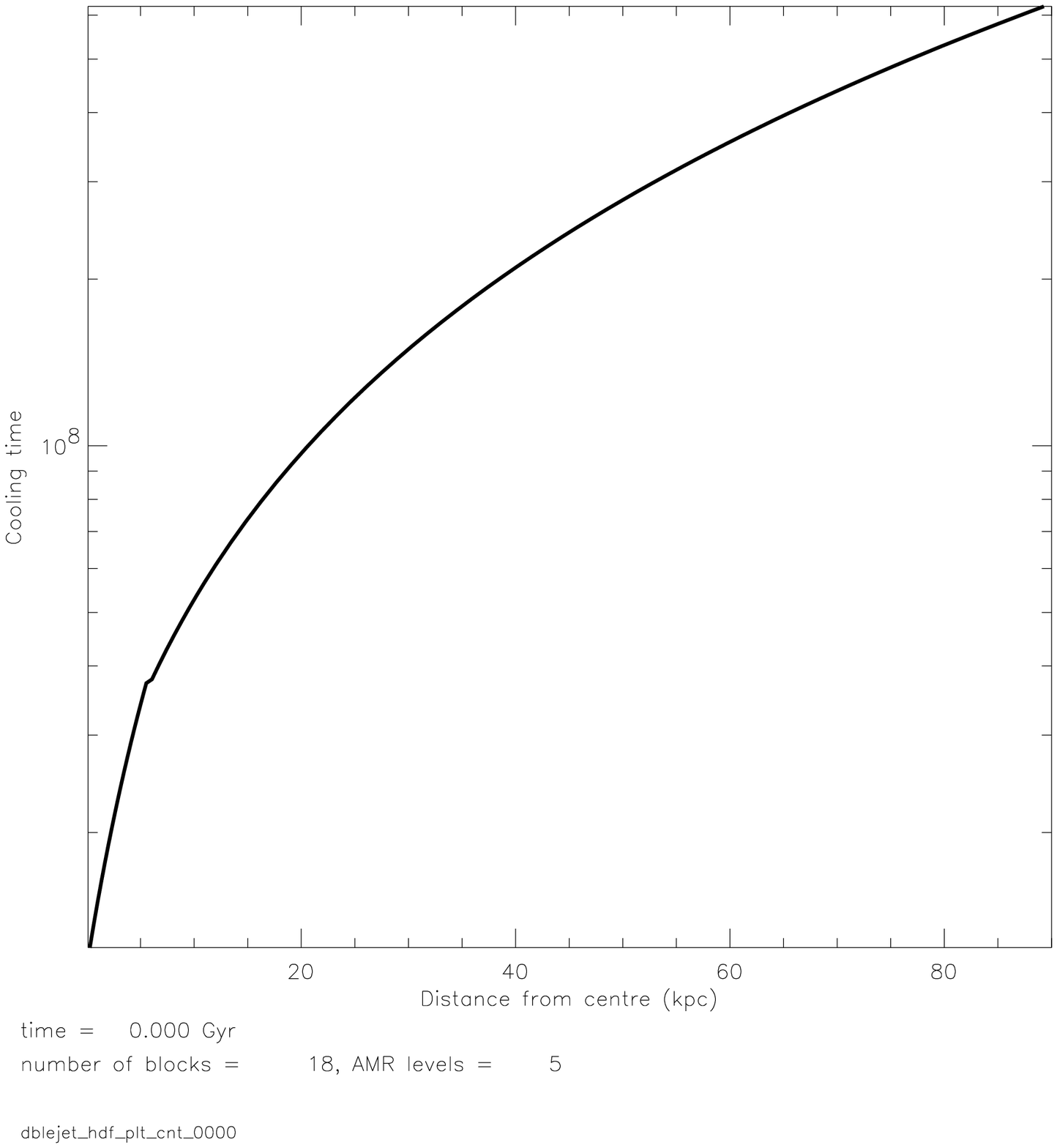}
\caption{Profiles of the initial density, temperature and cooling time.}
\label{init}
\end{figure}

The initial density was assumed to follow the dark matter, which is
given by a modified NFW profile (\citealt{nava97})
\begin{equation}
\rho(r)=\frac{\rho(0) r_s}{(r+r_s)(1+r/r_c)^2},
\end{equation}
where $r$ is the distance from the centre, $r_c$ the core radius and
$r_s$ a softening radius, which we assumed to be proportional to the
core radius, i.e. we chose $r_s=r_c/30$. The parameters of our model
are $n_{\rm e}(0)=0.3$ cm$^{-3}$, $r_c=300$ kpc, $r_s=r_c/30$ and the
baryon fraction is 10\%. The temperature and pressure are calculated
assuming initial hydrostatic equilibrium. The initial model of our
simulations is shown in Fig. \ref{init}. Thus our model has got a
central cooling time of around 20 Myrs.\\

Inspired by a growing number of observations of X-ray cavities in the
ICM, we introduce heating at off-centre sites by ad-hoc bubbles. To
mimic the energy input by a jet, hot, buoyant gas was injected
continuously into two small spherical regions which lie on opposite
sides of the centre of the cluster. The injection region had a radius
of 0.5 kpc, and the rate of energy input was $L=6\times10^{44}$ which
is a typical kinetic luminosity of a radio jet (\citealt{owen}). This rate
is also near the maximum rate that allows the bubbles to detach
themselves from the injection region in the course of their rise. For
higher luminosities a large bubble is inflated that evacuates most of
the cluster centre before it can rise and become detached from the
injection region. Which mechanism is responsible for tuning the energy
of the radio jets remains unclear. The energy input was switched off
after 208 Myr which is estimated to be a typical time for a recurring
FRI source (\citealt{sax01}). We assumed that the temperature of the
injected fluid was 100 times greater than the temperature of the
ambient fluid at the same location and that the gas had zero initial
velocity.\\

\begin{figure}[htp]
\plotone{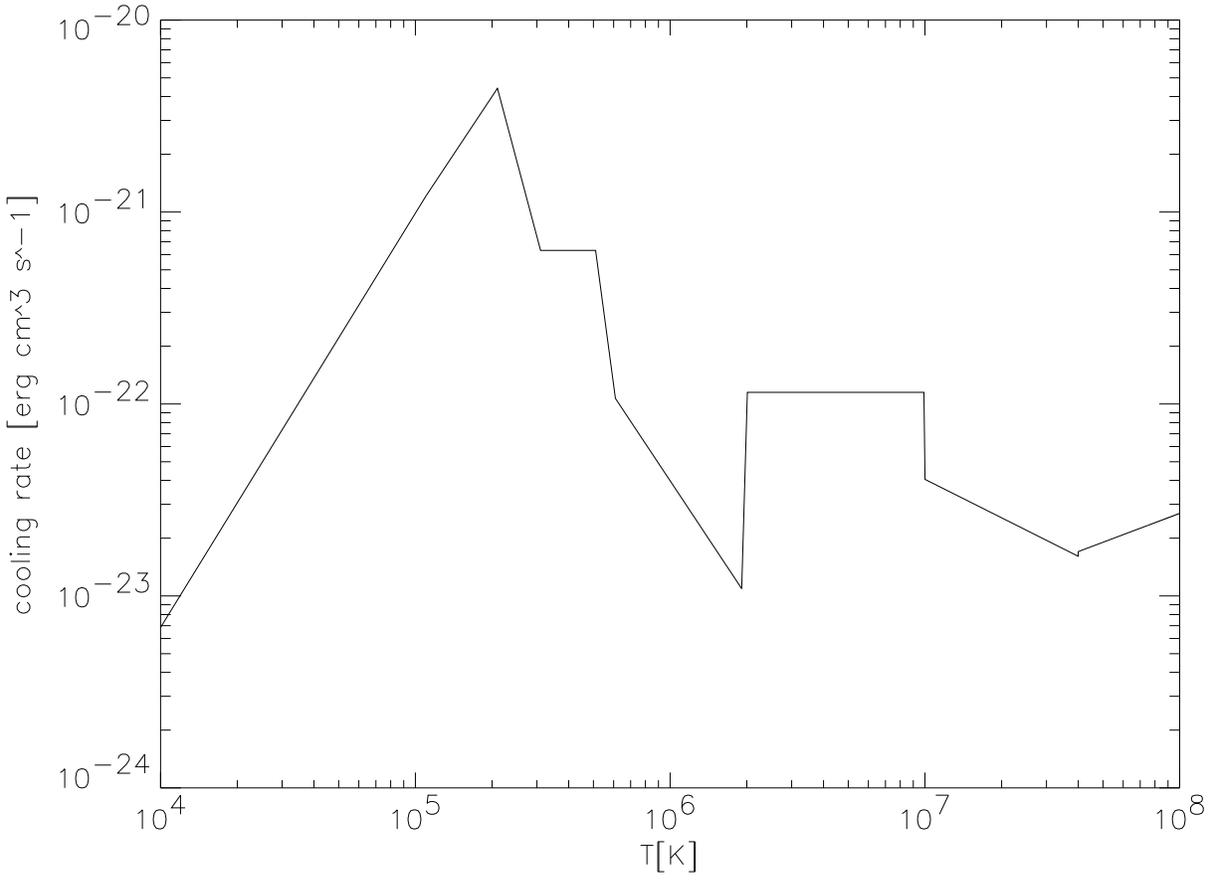}
\caption{Approximate cooling rate versus temperature as given by
Sarazin (1986) and Raymond (1976).}
\label{radia}
\end{figure}

We treated the hot gas in the bubbles and the ambient cluster gas as
two separate fluids which obey two separate equations of state. The
ICM was assumed to obey a $\gamma =5/3$ polytropic equation of
state. However, the equation of state within the bubbles is
unknown. We assume that the magnetic field within the rarefied bubbles
is tangled on small scales and can be approximated together with the
relativistic particles by a $\gamma =4/3$ equation of state. Thus, the
adiabatic index of the injected fluid in our simulation was $4/3$. The
ambient gas and the gas in the bubbles is allowed to cool via thermal
bremsstrahlung and line radiation. The cooling function of the gas was
taken from Sarazin (1986) and Raymond (1976) (see
Fig. \ref{radia}). We have neglected the synchrotron losses of the
radio plasma assuming it to be very hot so that it hardly cools at
all. Finally, we have produced purely hydrodynamic simulations and,
therefore, magnetic fields have been ignored. The only effect of
magnetic fields that has implicitly been taken into account is the
change of the adiabatic index of the injected fluid as described
above.\\

The simulations were performed with the FLASH code using an explicit
solver based on the piecewise-parabolic method (PPM). This method is
accurate to second order in both time and space and uses a
monotonicity constraint and contact steepening. For a detailed
description of the FLASH code, the reader is referred to
\cite{fryx}. Radiative cooling was treated explicitly with a
separate timestep limiter. The code uses an adaptive grid which
refines and derefines the computational mesh as needed. The
computational domain span 140 kpc $\times$ 140 kpc and was centred on
the centre of the cluster. The effective resolution in our simulations
was 2048 $\times$ 2048 zones. On all boundaries outflow conditions
were enforced. All simulations were run on 16 processors on an SGI
Origin 3000 and on an IBM Regatta system.

\section{Results and Discussion}

\begin{figure}[htp]
\plotone{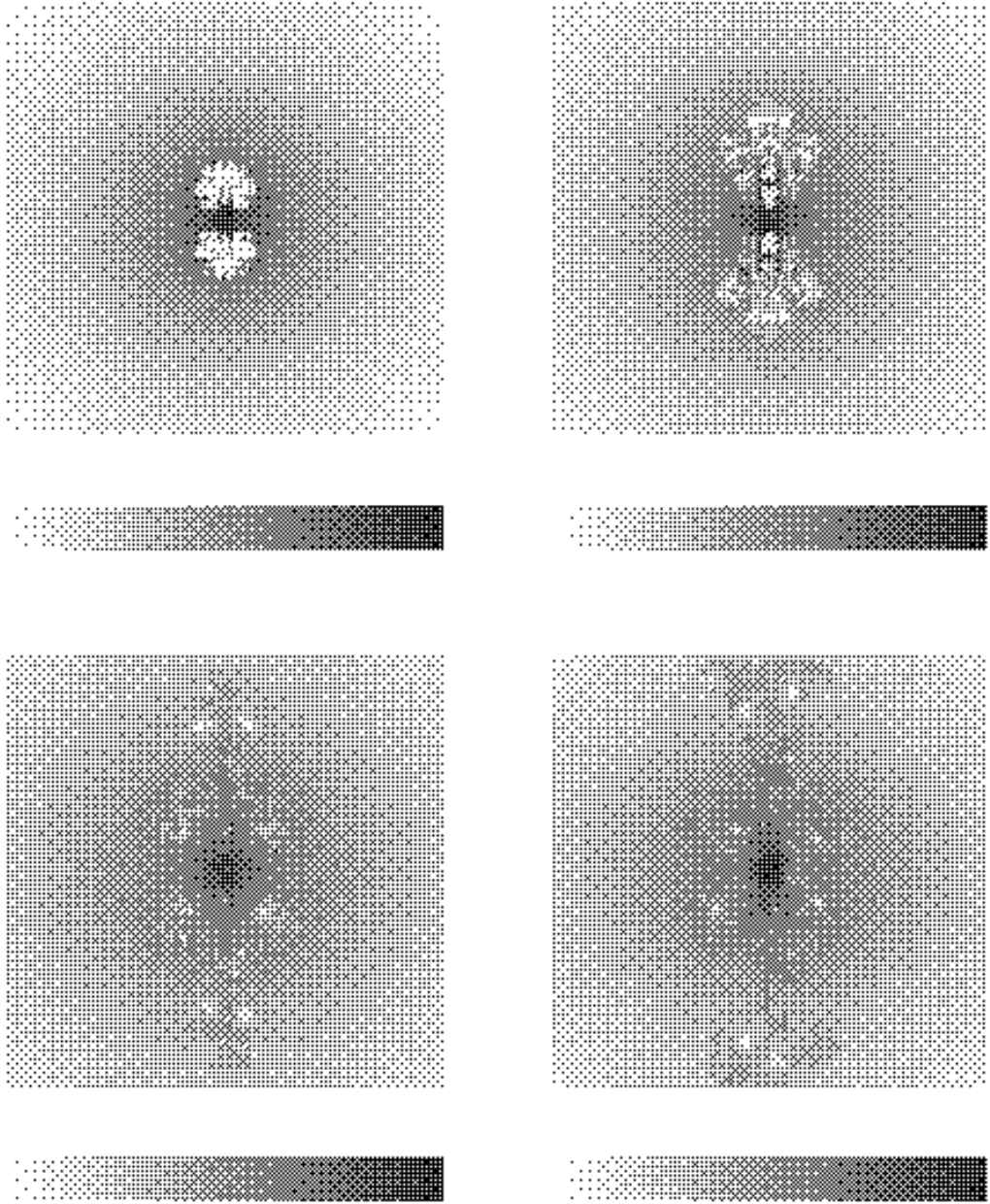}
\caption{Logarithmic contour plots of the density after 84, 167, 252 and 336 Myrs.}
\label{dens}
\end{figure}

Figure \ref{dens} shows logarithmic contour plots of the density at
84, 167, 252 and 336 Myr after the start of the energy injection. In
the last two pictures the injection of energy has ceased. One can see
how the buoyant plume rises upwards while Rayleigh-Taylor
instabilities form mushroom-shaped plumes on the bubble
surfaces. Kelvin-Helmholtz instabilities at the boundary between the
radio plasma and the ICM leads to the formation of tori and the
shedding of vortex rings. The tori form when the rarefied gas in the
wake of the bubble penetrates the bubble from below and makes contact
with the thermal gas on the upper side of the bubble
(\citealt{sax01}). The rising tori cause the thermal gas to circulate
around them. The velocity of the bubbles increase with increasing
distance from the centre as expected (\citealt{chu01}). A stream of
denser gas trails the bubbles and marks their path. The largest
bubbles are found at the leading edge of the buoyant gas. One can see
that as the bubble rises upwards the space between the top bubble and
the centre gets scattered with smaller radio bubbles and vortex
rings. Thus the bubbles are shredded producing a broad distribution of
sizes and velocities where the smaller bubbles rise more slowly. The
final fate of the bubbles, i.e. whether they will rise out of the
cluster before they are mixed microscopically into the ICM, is
difficult to tell from these simulations and will depend crucially on
the strength and structure of the magnetic fields inside the
bubbles. The simulations show that denser gas is pushed outwards
despite the effect of radiative cooling, which would cause the gas to
flow inward. It is remarkable that bubbles are found at fairly large
distances from the jet axis leading to a broad distribution of
bubbles. From an injection region of width of 1 kpc the bubbles spread
laterally to distances of up to 30 kpc - which corresponds to 1/10th
of the core radius - from the axis.

Thus the positions of X-ray cavities do not always allow simple
inferences on the orientation of the jet axis. In some cases bubbles
have been observed that are located off-centre from the current jet
axis and this has led to suggestions that the jet axis has
turned. However, this does not always have to be the case. Some
bubbles may merely have migrated away from the axis due to
instabilities. This may explain why in some clusters, such as Perseus,
the X-ray holes appear to be randomly distributed about the centre. \\

\begin{figure}[htp]
\plotone{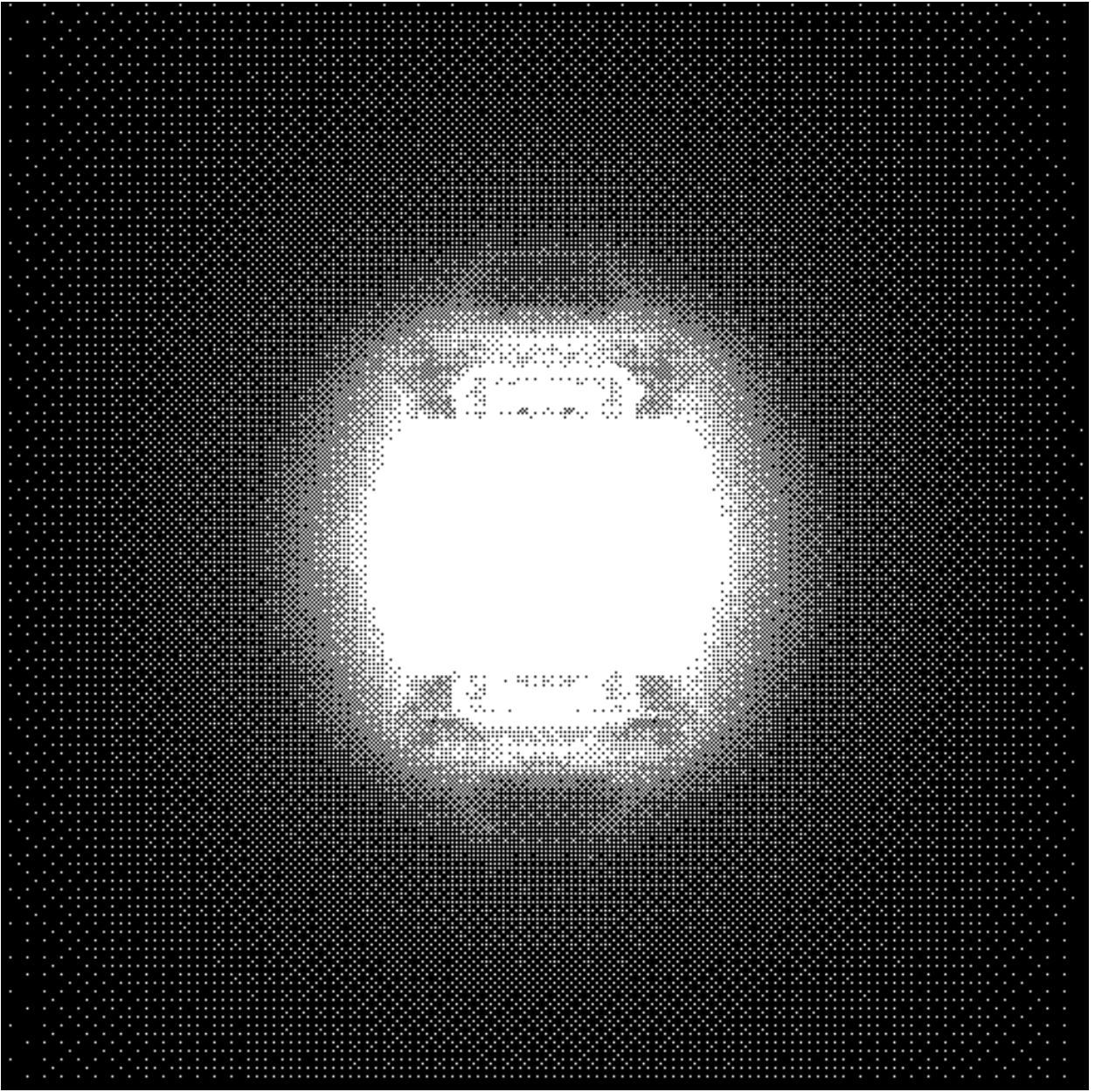}
\caption{Contour plots of the X-ray surface brightness after 100
Myrs. The image covers 140 $\times$ 140 kpc.}
\label{xray}
\end{figure}

In Fig. \ref{xray} we show a map of the X-ray surface brightness at a
time of 100 Myrs after the start of the simulation. For this purpose
we have rotated the 2D output around the jet axis and integrated the
bremsstrahlung emissivity along lines of sight. Since the
bremsstrahlung emissivity goes as $n_{\rm e}^2T^{1/2}$ the X-ray maps
show a very luminous core. Therefore, the cavities are more difficult
to observe at the centre but become more prominent once they have
reached a height of roughly a tenth of the core radius. The surface
brightness image also shows a bias towards larger bubbles while
many of the smaller structures remain obscured.\\

\begin{figure}[htp]
\plotone{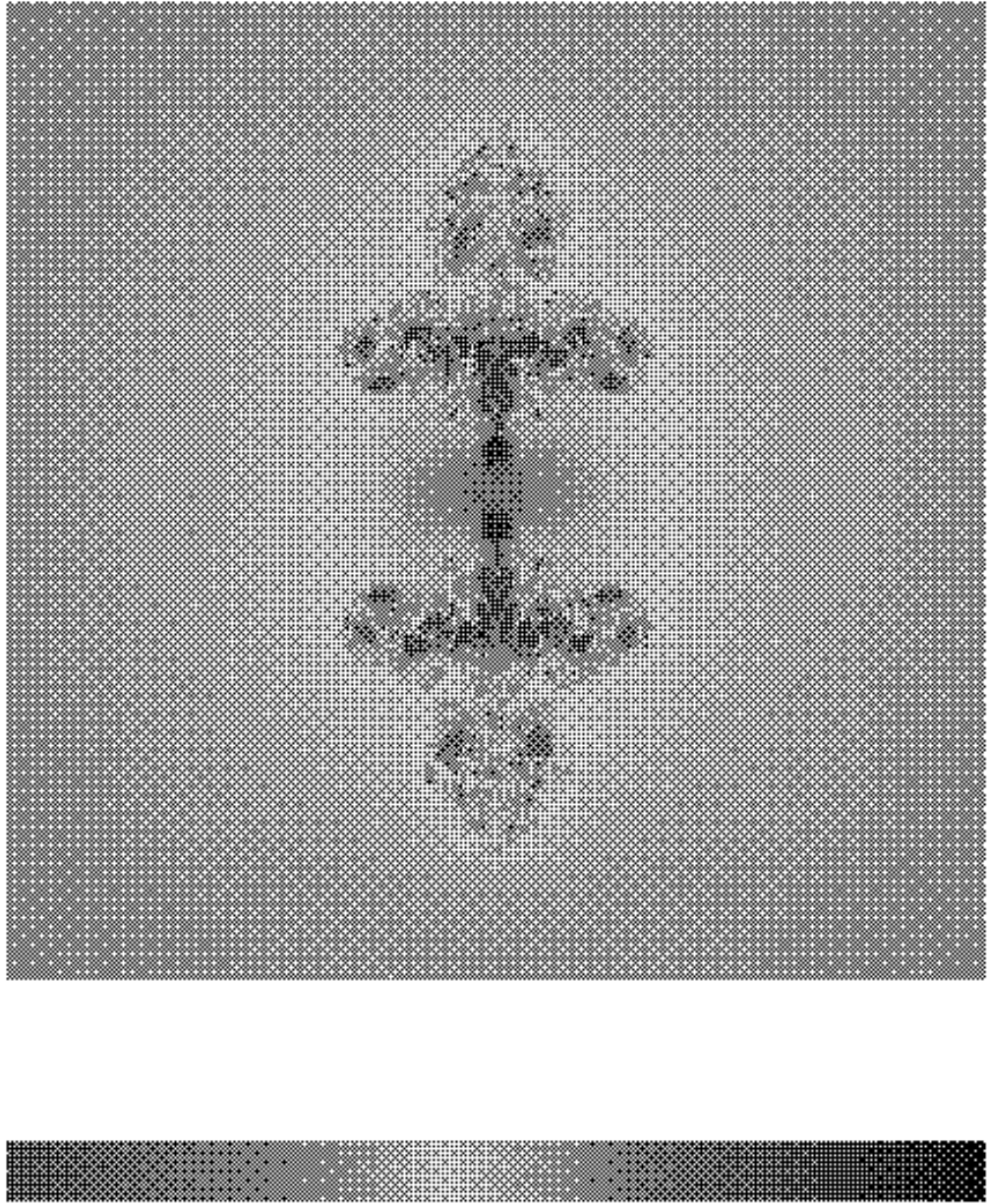}
\caption{Logarithmic contour plots of the temperature after 100
Myrs. The image covers 140 $\times$ 140 kpc.}
\label{temp}
\end{figure}

\begin{figure}[htp]
\plotone{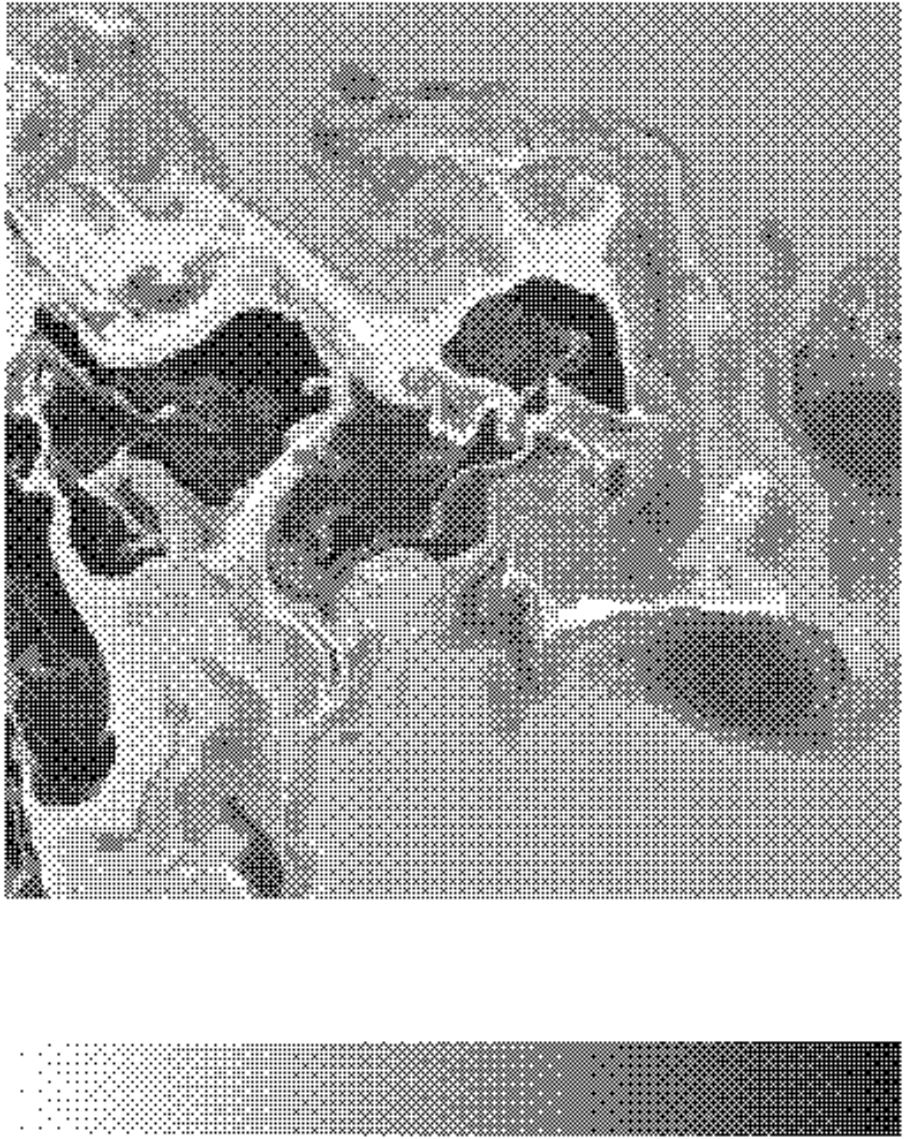} 
\caption{Close-up of the temperature around some buoyant bubbles after
100 Myrs. The colder regions around the bubbles correspond to the
bright rims seen in X-ray observations.}
\label{temp_zoom}
\end{figure}

Figure \ref{temp} shows a contour plot of the temperature at
120 Myr after the start of the energy injection. A trunk of colder gas
is uplifted in the wake of the bubbles. One can note how colder gas
has been entrained and is lifted upwards by the bubbles. As mentioned
in Sec. I, observations have shown that gas immediately around the
bubbles is colder than the ambient gas (e.g. Fabian et al. 2000, 2001,
\citealt{nul02}). In Fig. \ref{temp_zoom} we show an enlarged contour
plot of the temperature in the vicinity of the bubbles. One can
clearly see how entrained, colder gas from the central regions is
being pushed out by the buoyant bubbles. The bubbles are surrounded by
colder, low-entropy gas which appears bright in X-ray images. These
rims have got widths of $\sim$ 1 - 5 kpc. Their widths decrease with
increasing distance from the centre which can be attributed to three
factors: First, as the bubbles accelerate, they compress the gas in
front of them; second, the material in the rims cools faster than the
ambient ICM and third, some of the rim material is being mixed into
the ambient ICM. The rims are 20\% to 40\% cooler than the adjacent
thermal gas which is consistent with observations (\citealt{nul02}). They
also become cooler with their distance from the centre because they
cool faster.  Our simulations confirm the suggestion by Brighenti \&
Mathews (2002) who identified the source of the bright rims in
low-entropy gas from the cluster centre.

\begin{figure}[htp]
\plottwovert{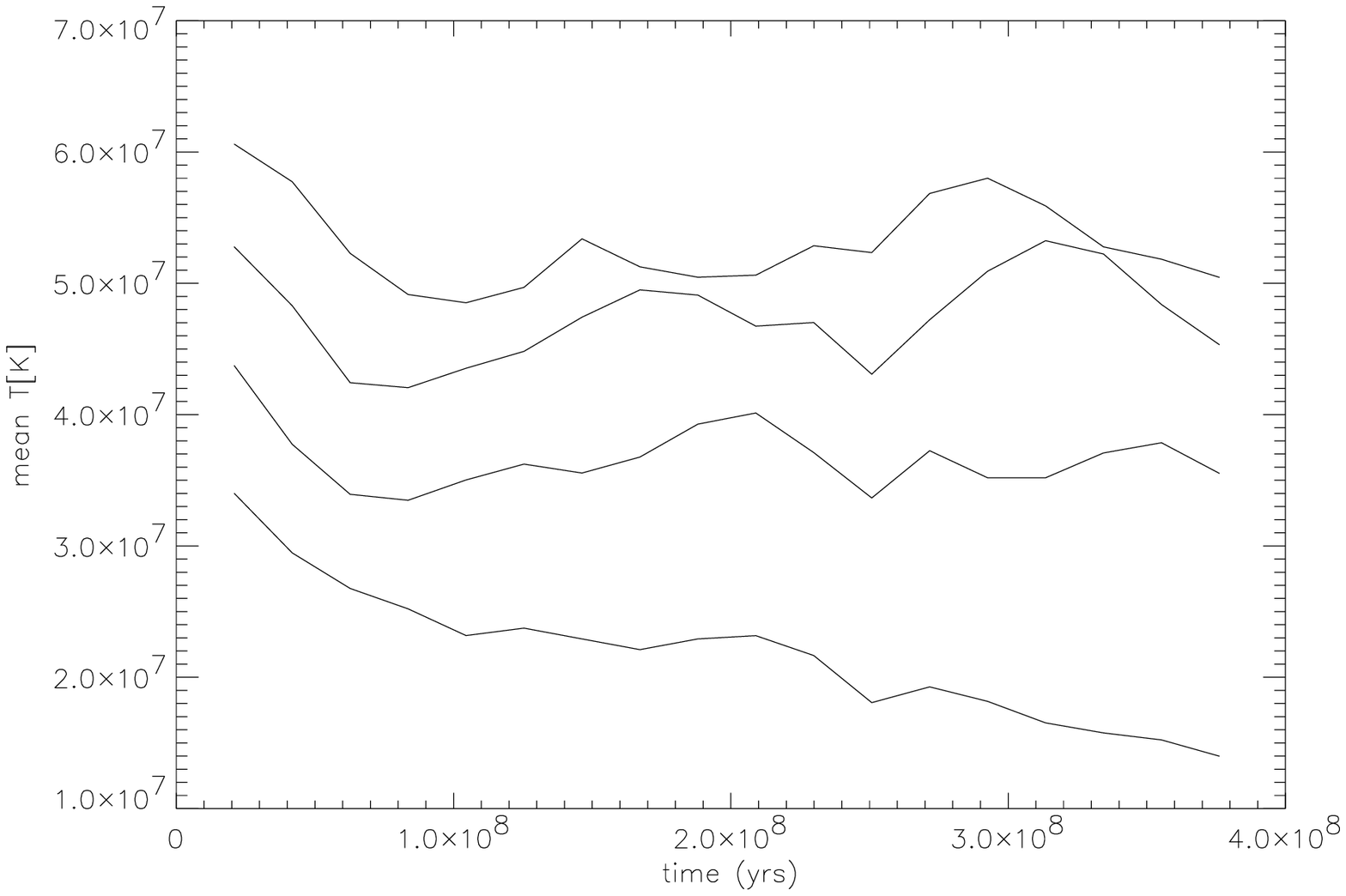}{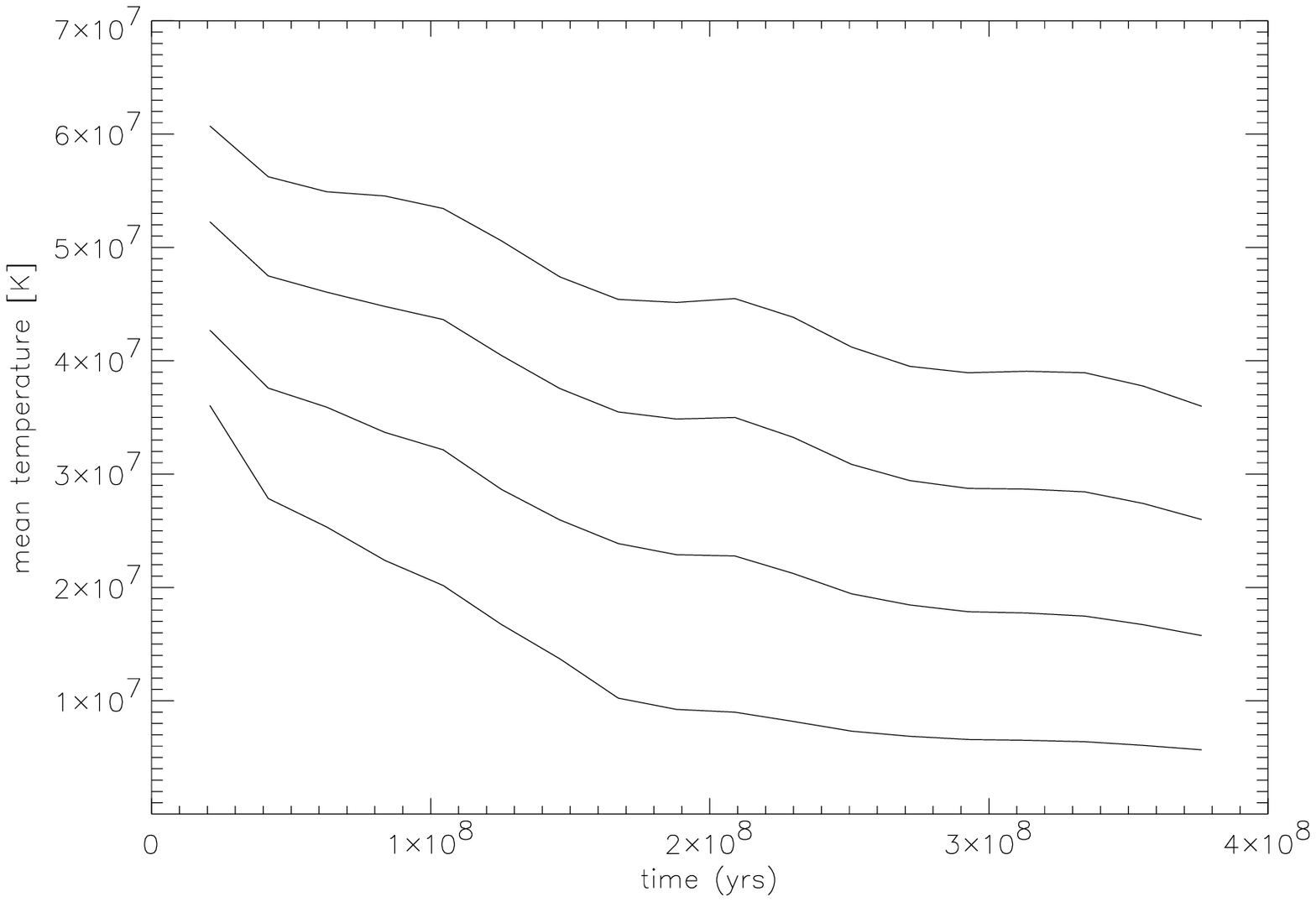}
\caption{Evolution of the temperature that has been averaged over
spherical shells about the centre. Each shell has a thickness of 5
kpc. Here the innermost 4 shells are shown which  going out to a maximum
radius of 20 kpc. The bottom curve corresponds to the innermost shell
with the temperature rising as we go farther out. The left panel shows
the average temperature in a simulation with heating by bubbles;
the right panel shows the average temperature for the identical cluster
without heating.}
\label{temp_evol}
\end{figure}
\begin{figure}[htp]
\plotone{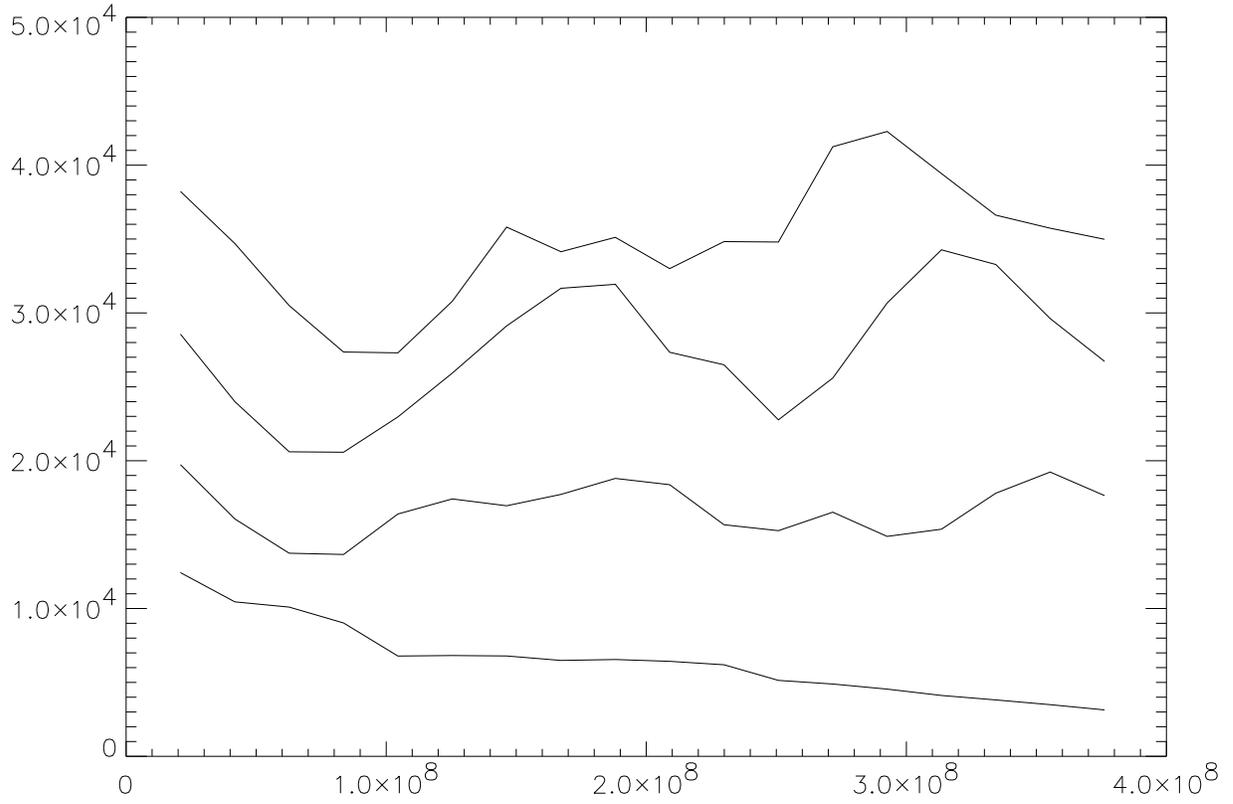} 
\caption{Evolution of the specific entropy averaged over spherical
shells about the centre. The x-axis shows the time in years.}
\label{entr_evol}
\end{figure}

We have found that without cooling the average temperature in the core
goes up because hotter material from outer regions is mixed into the
core and the expanding bubbles are performing work on the ambient
medium (\citealt{Bru02}). For the simulations presented here we have
plotted the temperature evolution of the cluster gas in
Fig. \ref{temp_evol}. The temperature has been averaged over spherical
shells of thickness 10 kpc about the cluster centre. The code treats
the ambient and the injected material as two separate fluids
throughout, which enables us to study the temperature evolution of the
original cluster gas alone. Only the ambient cluster gas has been
included in the average shown in Fig. \ref{temp_evol}. In a separate
run, we have switched off the energy input by the bubbles and the
corresponding temperature evolution is shown in the right panel of
Fig. \ref{temp_evol}.  The comparison shows clearly how, over the time
scales considered here, the bubbles manage to keep the temperature in
the central regions of the cluster constant (or even raise them),
whereas in the run without energy input we see the usual
radiatively-regulated cooling inflow. Fig. \ref{entr_evol} shows the
evolution of the specific entropy in the cluster. Again, the entropy
has been averaged over spherical shells of thickness 5 kpc. One can
notice how in all but the innermost shells the entropy is kept at a
fairly constant level or even rises despite the cooling. This effect
caused by the rising bubbles persists for a time of more than 400 Myrs
which is twice as long as the time for which the jet was active and
approximately the sound crossing time of the core radius $r_c$. In the
cooling-flow model, mass deposition is greatest in the core, where the
cooling time drops sharply. The mass that condenses out in the cores
ceases to provide pressure support, and thus causes a slow
inflow. Therefore, the cooling flow can be delayed significantly if
the cooling of gas in the central regions can be halted. In order to
gauge what fraction of the energy of the injected material is
transferred to the ICM, we have repeated our simulation with the
cooling switched off. We then calculate the total internal energy of
the ambient gas and find that after 200 Myrs $\sim 10$ \% of the
injected energy has been deposited in the cluster gas. This
calculation ignores that the injection of hot material might cause
enhanced cooling in some of the cluster gas but we estimate that this
effect is not very large.\\

One can conceive of various physical mechanisms whereby the energy of
the hot and buoyant fluid is transferred to the ICM: (i) $pdV$ work on
the ICM by the expanding bubbles (ii) entrainment of ambient fluid,
(iii) microscopic mixing between bubble and ambient fluid, (iv)
turbulent excitation of sound and gravity waves and (v) thermal
conduction.

As pointed out in \citet{chu02} the work done by the rising bubble can
be estimated by considering the enthalpy of the bubble during
adiabatic expansion

\begin{equation}
H=\frac{\gamma}{\gamma-1}pV=H_0\left ( \frac{p}{p_0}\right) \ ,
\end{equation}
where $H_0$ and $p_0$ are the enthalpy and pressure at the initial
position. Taking a typical pressure profile, approximately half the
energy is lost at a distance of $\sim$ 20 kpc from the origin
(\citealt{chu02}). This work goes mainly into $pdV$ work on the
ambient medium and kinetic and potential energy of the entrained
gas. Given the steep entropy profile in the inner regions of the
cluster, large masses of entrained gas cannot travel a long distance
with the rising bubbles, and therefore most of the energy will stay in
the cluster.

Microscopic mixing is not included in our code and any small-scale
mixing that occurs in the simulation must therefore be entirely
numerical. Convergence tests with our code have shown that any
numerical mixing is negligible.

Sound waves will only have a minor share in the energy lost since the
ratio between the energy in sound waves over the dissipated energy for
subsonic turbulence goes approximately as (\citealt{ll})

\begin{equation}
\frac{\epsilon_s}{\epsilon_d} \sim \left ( \frac{v}{c_s}\right)^5 \ll 1 \ .
\end{equation}
Since most of the turbulent motions induced by the buoyant bubbles are
subsonic, this fraction is going to be small. Estimates made in
\citet{chu02} show that the energy emitted in sound waves constitutes
not more than a small fraction of the internal energy of the bubble.

Finally, thermal conduction is not included in our simulations. The
role of thermal conduction in clusters is a matter of contention, with
most observations pointing towards a heavily suppressed thermal
conductivity in the ICM (e.g. \citealt{fabi94}). Hence, we conclude
that $pdV$ work by the rising and expanding bubbles as well as
turbulent entrainment are the most likely mechanisms whereby the radio
plasma deposits energy in the ICM.

Clearly, the lifetime of the activity of the central AGN is short
compared to the evolutionary time scale of the cluster gas. Therefore,
once the AGN stops supplying energy to the buoyant bubbles, the
cluster gas will settle down once again and a full cooling flow may be
re-established. The cooling gas flowing to the centre of the cluster
may then trigger a further active phase of the AGN. Thus a
self-regulating process with cooling periods alternating with brief
bursts of AGN activity may be established. The rising bubbles uplift
colder material from the vicinity of the AGN and thus disrupt the
supply of fresh fuel for the radio jets. This feedback mechanism might
automatically regulate the power of the radio jets. The details of
this mechanism remain unclear but it is potentially important as it
might provide an upper mass limit for galaxies. For more on feedback
between gas accretion and radio power see, e.g. \cite{qui01},
\cite{voit01}, \cite{ruszkow}, Churazov et al. (2002b).\\

It has been found that 71 \% of all cD galaxies at the centres of
cooling-flow clusters show evidence of radio activity
(\citealt{bur90}). This fraction is higher than in non-cooling flow
galaxies which again points towards the existence of some form of
feedback mechanism. If all AGN at the centres of cooling flows go
through activity cycles of constant length, $t_{\rm on}$, separated by
periods of quiescence, again of constant length, $t_{\rm off}$, then
these observations imply that $t_{\rm off} < 0.41\ t_{\rm on}$. Figure
\ref{entr_evol} indicates that the simulated cluster starts to relax
back to its original state after a time of $\sim 400$ Myr and returns
to a cooling dominated inflow. Thus it gets close to a configuration
similar to that at which the AGN activity started. Despite having
chosen one particular realisation of the gas distribution inside the
cluster, these findings are insensitive to the details of the cluster
model. Experiments with other cluster models have yielded very similar
results. Here we have only presented results of 2D simulations because
we wanted to achive the highest effective resolution that we could
afford. However, previous 3D simulations (\citealt{Br}) of lower
resolution have shown that the gross features and integrated
quantities do not differ greatly from 2D simulations.\\
 
Cosmological simulations as well as observations suggest that the gas
in many clusters may not be in an equilibrium state, in contrast to
what we assumed in our simulation. However, gas flows of low Mach
numbers within a cluster are unable to impede the rise of the bubbles,
and are more likely to increase than to decrease the mixing. Moreover,
clusters that are far from hydrostatic equilibrium are generally not
expected to host cooling flows.\\

In summary, we have found that bubbles can heat the cluster core by
input of mechanical energy. The effect of the bubbles on the ICM
consists of heating via $PdV$ work and redistribution of mass via
buoyancy-driven mixing.  For realistic parameters the heating is shown
to be sufficient to balance radiative cooling and to disrupt, at least
temporarily, the cooling flow. Thus a significant fraction ($\sim 10$
\%) of the energy residing in radio lobes can be dissipated in the
cooling flow. Clearly, heating by AGN will not account for the
entirety of the missing gas in cooling flows but it is shown to be an
efficient and common mechanism which introduces heat into the ICM.
During their rise in the cluster potential, the bubbles expand and
produce depressions in the X-ray surface brightness. Moreover,
hydrodynamical instabilities produce a broad distribution of smaller
bubbles and vortices. Bright rims are a common feature around buoyant
bubbles and consist of colder gas that has been uplifted from the
centre. The bright rims seen in our simulations are consistent with
those found in observations.

\acknowledgments The software used in this work was in part developed
by the DOE supported ASCI/Alliances Center for Thermonuclear Flashes
at the University of Chicago. I thank Christian Kaiser, Bill Forman
and Eugene Churazov for helpful discussions.

\clearpage


\begin{thebibliography}{}

\bibitem[Basson \& Alexander(2002)]{Bas02}Basson, J.,\& Alexander,
P. 2002, submitted to MNRAS

\bibitem[Bertschinger \& Meiksin(1986)]{bert86}Bertschinger, E., \&
Meiksin, A. 1986, ApJ, 306, L1

\bibitem[Bicknell \& Begelman(1996)]{bic96} Bicknell, 
G.~V.~\& Begelman, M.~C.\ 1996, \apj, 467, 597 

\bibitem[Binney(1995)]{1995} Binney, J. \& Tabor, G. 1995, MNRAS, 276, 663

\bibitem[Blanton et al.(2001)]{blan01}Blanton, E.~L., Sarazin, C.~L.,
McNamara, B.~R., \& Wise, M.~W. 2001, \apj, 558, L15

\bibitem[B\"ohringer et al.(1993)]{boe93} B\"ohringer, H., 
Voges, W., Fabian, A.~C., Edge, A.~C., \& Neumann, D.~M.\ 1993, \mnras, 
264, L25 

\bibitem[B{\" o}hringer et al.(2002)]{boe02} B{\" o}hringer, 
H., Matsushita, K., Churazov, E., Ikebe, Y., \& Chen, Y.\ 2002, \aap, 382, 
804

\bibitem[Brighenti \& Mathews(2002)]{2002ApJ...573..542B} Brighenti, F.~\& 
Mathews, W.~G.\ 2002, \apj, 573, 542

\bibitem[Br{\" u}ggen \& Kaiser(2002)]{Bru02} Br{\" u}ggen, 
M.~\& Kaiser, C.~R.\ 2002, \nat, 418, 301 

\bibitem[Br{\" u}ggen et al.(2002)]{Br} Br{\" u}ggen, M., Kaiser, C.~R.\
R., Churazov, E., Ensslin, T.~A.. 2002, MNRAS, \mnras, 331, 545

\bibitem[Burns(1990)]{bur90} Burns, J.~O.\ 1990, \aj, 99, 14 

\bibitem[Carilli, Perley, \& Harris(1994)]{car94} Carilli, 
C.~L., Perley, R.~A., \& Harris, D.~E.\ 1994, \mnras, 270, 173 

\bibitem[Churazov, Forman, Jones, \& B{\" 
o}hringer(2000)]{2000A&A...356..788C} Churazov, E., Forman, W., Jones, C., 
\& B{\" o}hringer, H.\ 2000, \aap, 356, 788 

\bibitem[Churazov et al.(2001)]{chu01} Churazov, E. and Br{\" u}ggen,
M. and Kaiser, C.\ R. and B{\" o}hringer, H. and Forman, W.\ 2001a,
\apj, 554, 261

\bibitem[Churazov, Sunyaev, Forman, \& B{\" 
o}hringer(2002b)]{2002MNRAS.332..729C} Churazov, E., Sunyaev, R., Forman, 
W., \& B{\" o}hringer, H.\ 2002, \mnras, 332, 729 

\bibitem[Churazov et al.(2002)]{chu02} Churazov, E., B{\" 
o}hringer, H., Br{\" u}ggen, M., Forman, W., Jones, C., Kaiser, C., \& 
Sunyaev, R.\ 2002, Lighthouses of the Universe: The Most Luminous Celestial 
Objects and Their Use for Cosmology Proceedings of the MPA/ESO/, p.~37, 37 

\bibitem[Cowie \& Binney(1977)]{1977ApJ...215..723C} Cowie, L.~L.,
Binney, J.\ 1977, \apj, 215, 723

\bibitem[Crawford et al.(1999)]{craw99} Crawford, C.~S., 
Allen, S.~W., Ebeling, H., Edge, A.~C., \& Fabian, A.~C.\ 1999, \mnras, 
306, 857 

\bibitem[Fabian \& Nulsen(1977)]{1977MNRAS.180..479F} Fabian, A.~C.~\& 
Nulsen, P.~E.~J.\ 1977, \mnras, 180, 479 

\bibitem[Fabian (1994)]{fabi94}Fabian, A.C. 1994, \araa, 32, 277

\bibitem[Fabian et al.(2000)]{f00} 
Fabian A.~C., Sanders J.~S., Ettori S., Taylor G.~B., Allen S.~W.,
Crawford C.~S., Iwasawa K., Johnstone R.~M., Ogle P.~M., 2000, \mnras,
accepted, astro-ph/0007456

\bibitem[Fabian et al.(2002)]{fabi02}Fabian, A.~C., Voigt, L.~M., \&
Morris, R.~G., 2002, \mnras, submitted (astro-ph/0206437)

\bibitem[Fryxell et al.(2000)]{fryx} Fryxell, B.~et al.\ 
2000, \apjs, 131, 273

\bibitem[Gruzinov(2002)]{gruz02}Gruzinov, A. 2002, astro-ph/0203031

\bibitem[Landau \& Lifshitz(1987)]{ll} L.~D. Landau, L.D. \& Lifshitz,
E.~M., Fluid Mechanics. Pergamon, Oxford, 2nd edition, 1987

\bibitem[Mazzotta et al.(2002)]{mazz02} Mazzotta, P., Kaastra, 
J.~S., Paerels, F.~B., Ferrigno, C., Colafrancesco, S., Mewe, R., \& 
Forman, W.~R.\ 2002, \apjl, 567, L37 

\bibitem[McNamara(1997)]{mcnam97} McNamara, B.~R.\ 1997, ASP 
Conf.~Ser.~115: Galactic Cluster Cooling Flows, 109 

\bibitem[McNamara et al.(2001)]{mcnam01} McNamara, B.~R.~et 
al.\ 2001, \apjl, 562, L149 

\bibitem[McNamara(2002)]{mcnam02} McNamara, B.~R., 2002, Invited review,
"The High-Energy Universe at Sharp Focus: Chandra Science," Symposium
at the ASP meeting, 16-18 July, 2001, St. Paul, MN

\bibitem[Narayan \& Medvedev(2001)]{nara01}Narayan, R., \& Medvedev,
M.V. 2001, \apj, 562, L129

\bibitem[Navarro et al.(1997)]{nava97}Navarro, J.F., Frenk, C.S., \&
White, S.D.M. 1997, \apj, 490, 493

\bibitem[Nulsen et al.(2002)]{nul02} Nulsen, P.~E.~J., David, 
L.~P., McNamara, B.~R., Jones, C., Forman, W.~R., \& Wise, M.\ 2002, \apj, 
568, 163 

\bibitem[Owen, Eilek, \& Kassim(2000)]{owen} Owen, F.~N., 
Eilek, J.~A., \& Kassim, N.~E.\ 2000, \apj, 543, 611 

\bibitem[Peterson et al.(2001)]{pete01}Peterson, J.~R., Paerels, F.B.S.,
Kaastra, J.~S., et al. 2001, \aap, 365, L104

\bibitem[Quilis, Bower, \& Balogh(2001)]{qui01} Quilis, V., 
Bower, R.~G., \& Balogh, M.~L.\ 2001, \mnras, 328, 1091 

\bibitem[Raymond, Cox, \& Smith(1976)]{1976ApJ...204..290R} Raymond, J.~C., 
Cox, D.~P., \& Smith, B.~W.\ 1976, \apj, 204, 290 

\bibitem[Reynolds, Heinz, \& Begelman(2001)]{RHB01} Reynolds, 
C.~S., Heinz, S., \& Begelman, M.~C.\ 2001, \apjl, 549, L179 

\bibitem[Ruszkowski \& Begelman (2002)]{ruszkow} Ruszkowski, M., \& Begelman, M. 2002, \apj, 581, 223

\bibitem[Sarazin(1986)]{1986RvMP...58....1S} Sarazin, C.~L.\ 1986, Reviews 
of Modern Physics, 58, 1 

\bibitem[Sarazin(1988)]{1988cfcg.work....1S} Sarazin, C.~L.\ 1988,
NATO ASIC Proc.~229: Cooling Flows in Clusters and Galaxies, Kluwer,
Dordrecht, 1

\bibitem[Saxton, Sutherland, \& Bicknell(2001)]{sax01} 
Saxton, C.~J., Sutherland, R.~S., \& Bicknell, G.~V.\ 2001, \apj, 563, 103 

\bibitem[Schindler et al.(2001)]{sch01} Schindler, S., 
Castillo-Morales, A., De Filippis, E., Schwope, A., \& Wambsganss, J.\ 
2001, \aap, 376, L27 

\bibitem[Soker \& Sarazin(1990)]{sok90} Soker, N.~\& Sarazin, 
C.~L.\ 1990, \apj, 348, 73 

\bibitem[Tabor \& Binney(1993)]{tabo93}Tabor, G., \& Binney, J. 1993,
\mnras, 263, 323

\bibitem[Tamura et al.(2001)]{tamu01}Tamura, T., Kaastra, J.~S.,
Peterson, J.~R., Paerels, F., et al. 2001, \aap, 365, L87

\bibitem[Voigt et al.(2002)]{voig02}Voigt, L.M., Schmidt, R.W.,
Fabian, A.C., Allen, S.W., \& Johnstone R.M., 2002, \mnras, submitted
(astro-ph/0203312)

\bibitem[Voit \& Bryan(2001)]{voit01} Voit, G.~M.~\& Bryan, 
G.~L.\ 2001, \nat, 414, 425 

\bibitem[Vrtilek et al.(2000)]{vrt} Vrtilek, J.~M., David, 
L.~P., Grego, L., Jerius, D., Jones, C., Forman, W., Donnelly, R.~H., \& 
Ponman, T.~J.\ 2000, Constructing the Universe with Clusters of Galaxies,

\bibitem[Wilson, Young, \& Shopbell(2000)]{wil00} Wilson, 
A.~S., Young, A.~J., \& Shopbell, P.~L.\ 2000, \apjl, 544, L27 

\bibitem[Zakamska \& Narayan (2002)]{zaknar02} Zakamska, N.L. \&
Narayan, R., 2002, \apj, in press

\end{thebibliography}
\end{document}